\documentclass[natbib]{aastex}
\usepackage{spr-astr-addons}
\usepackage{url}
\usepackage{graphicx}
\usepackage{amsmath}
\usepackage{amssymb}
\usepackage{lscape}
\usepackage{color} 
\usepackage{booktabs}
\usepackage{bm}
\usepackage{multirow}
\usepackage{array}
\usepackage{adjustbox}

\RequirePackage{color}

\newfont{\myfont}{cmmib10}

\newcommand{\bphi}{\hbox{\myfont \symbol{30} }}
\newcommand{\bomega}{\hbox{\myfont \symbol{33} }}
\newcommand{\btheta}{\hbox{\myfont \symbol{18} }}

\def\lapprox{\mathrel{\hbox{\rlap{\hbox{\lower4pt\hbox{$\sim$}}}\hbox{$<$}}}}
\def\gapprox{\mathrel{\hbox{\rlap{\hbox{\lower3pt\hbox{$\sim$}}}\hbox{$>$}}}}

\begin{document}

\title{Properties of the emission region in pulsars with opposite subpulse drift directions in different profile components}
\slugcomment{}
\shorttitle{Properties of the emission region in pulsars}
\shortauthors{R. Yuen et al.}

\author{H. M. Tedila\altaffilmark{1,2,3}}
\and
\author{R. Yuen\altaffilmark{$^\dagger$1,4,5}}
\and
\author{X. H. Han\altaffilmark{1}}
\altaffiltext{$^\dagger$}{E-mail: ryuen@xao.ac.cn}
\altaffiltext{1}{Xinjiang Astronomical Observatory, Chinese Academy of Sciences, 150 Science 1-Street, Urumqi, 
Xinjiang 830011, China}
\altaffiltext{2}{University of Chinese Academy of Sciences, 19A Yuquan Road, 100049 Beijing, PR China}
\altaffiltext{3}{Arba Minch University, Arba Minch 21, Ethiopia}
\altaffiltext{4}{Key Laboratory of Radio Astronomy, Chinese Academy of Sciences, 150 Science 1-Street, Urumqi, Xinjiang 830011, China}
\altaffiltext{5}{SIfA, School of Physics, University of Sydney, Sydney, NSW 2006, Australia}
\email{\yuen@xao.ac.cn}

\begin{abstract}
We investigate properties of the emission region as revealed by drifting subpulses of opposite drift directions at different parts of a pulse profile by using the rotating carousel model in an obliquely rotating pulsar magnetosphere of multiple emission states. Subpulse emission is assumed coming from $m$ discrete emission areas that are distributed around the magnetic axis on a rotating carousel. The flow rate of the emission areas is determined by the $\bm{E}\times \bm{B}$ drift in an emission state, designated by the parameter $y$, in which $\bm{E}$ and the associated flow rate are dependent on $y$. In this model, subpulses appear to drift in an emission state if a relative speed exists between the plasma flow and corotation, and the diversity in the drift rates and directions corresponds to the relative speed being different in different parts of a profile. We apply the model to three pulsars that exhibit drifting subpulses of opposite drift directions to identify the emission states and the values of $m$. Our results show that different drifting subpulses correspond to particular values of $m$ and $y$, and the latter implies that different emission states can coexist and operate concurrently in an emission region. We find that $m$ does not show clear dependency on either the obliquity angle or emission state. We demonstrate that subpulse arrangement may vary across an emission region meaning that it is not always uniform on a carousel. We discuss drifting subpulses of opposite drift directions and subpulse drift-rate switching in terms of different emission states in our model, and speculate that they may be two manifestations of the same underlying mechanism.

\end{abstract}

\keywords{pulsars: general -- stars: neutron -- radiation mechanisms: non-thermal}

\section{Introduction}

Observations of single pulses in some radio pulsars reveal intriguing features displaying as drifting subpulses of opposite drift directions along drift-bands that are located across different parts of the integrated pulse profile \citep{MLC+04, Wetal06, Wetal07, Weltevrede16}. Perhaps an extreme example is that found in PSR J0815+0939 \citep{MLC+04, CLM+05}. Termed as bi-drifting, the subpulses in the pulsar exhibit drifting along four different drift-bands located in different profile components with the drift direction of the drift-band in the second component being opposite to that of the others \citep{SL17}. The phenomenon poses a challenge to the traditional understanding of drifting subpulses, and provides a unique way to explore pulsar radio emission \citep{QLZ+04, SL17}. The traditional explanation for drifting subpulses, the systematic marching of subpulses across a profile in sequence of pulses, usually involves confining the emitting locations to discrete subbeams. The subbeams are placed evenly around the magnetic axis on a rotating carousel under the $\bm{E} \times \bm{B}$ drift \citep{Ruderman72, RudermanSutherland1975, DR99, DR01, ES02}. In this rotating carousel model, the subbeams rotate relative to corotation through the fixed line of sight causing the subpulses to drift. The model is successful in explaining the fundamental characteristics of the phenomenon, from which basic pulsar parameters, such as the viewing angle, $\zeta$, between the line of sight and the rotation axis, and the obliquity angle, $\alpha$, between the magnetic and the rotation axes, can be inferred. Furthermore, the assumption of a well-organized subpulses in singular emission state leads to uni-directional drift. Therefore, drifting subpulses with opposite drift directions in different profile components, abbreviated here as opposite subpulse drifting, suggest that modification is needed for the carousel model. This is because when the opposite subpulse drifting along different drift-bands are considered separately, they appear ``normal'' in the sense that each can be described using conventional drift parameters based on the carousel model. However, since subpulse drifting is strongly linked to emission properties in the emission region \citep{RudermanSutherland1975}, opposite subpulse drifting also suggests that multiple emission states, each with different emission properties, can coexist in the emission region. 

Coexistence of different emission states has been observed in several radio pulsars. A classic example relates to the episodes of nulling (disappearance of pulses) in intermittent pulsars \citep{KLO+06, CRC+12, LLM+12, LSF+17, WWH+20}. The cyclical change in emission between `on' (pulse detection) and `off' (pulse non-detection) is proposed to involve switching between two emission states of charge-filled and vacuum in the magnetosphere \citep{KLO+06}. This implies that the entire emission region is occupied with a similar state at any one time and changes take place concurrently over the entire region. In this case, the existence of different emission states is exclusive from one another and the effects of each manifest at different times. Unique emission properties can also be associated with different ranges of longitudinal phase. The observation of pulse disappearance in PSR J1819+1305 from only the trailing component accompanying with changes in the drift pattern in the two leading components \citep{RW08} imply that different emission states are operating at the same time across different parts of the pulse profile. Another similar example is that found in PSR B2020+28 which displays different nulling fractions between the leading and trailing components \citep{GJK12}. In these pulsars, different emission states can coexist and function concurrently with the effects of each being confined to a specific range of longitudinal phase within the profile. However, how different emission states can coexist in an emission region, and how the emission properties are different in different emission states that give rise to the phenomenon remain unclear. In the case of opposite subpulse drifting, the singular emission state implied in the carousel model is insufficient for identifying the different emission properties in different emission states in relation to the phenomenon. Furthermore, the assumption that subpulses drift because of the flow rate of the emitting plasma deviates from corotation implies that the plasma flow rate is also dependent on the emission state. This means that corotation is merely one case of plasma flow, which results in zero drift rate. Hence, subpulse drifting and non-drifting are likely two manifestations of the same mechanism. This makes studying opposite subpulse drifting important as it provides additional insights into the emission region and how it relates to drifting subpulses in the extreme case.

The purpose of this paper is to explore the emission properties and the different emission states as revealed by opposite subpulse drifting in different pulsars. Our analysis is based on the rotating carousel incorporating the model for pulsar magnetospheres of multiple emission states, designated by the parameter $y$ \citep{MY14a, MY16, Yuen19b}. We assume a periodic structure of overdense (subbeams) and underdense regions of plasma that varies in proportional to $\cos(m\phi_b)$ \citep{CR04, GMM+05}, where $m$ is an integer, in azimuthal direction around the magnetic axis as a result of a standing wave at a specific spherical harmonic determined by an instability in the magnetosphere \citep{FKK06, Petri07}. The subpulse drift velocity is the flow rate of the emitting plasma due to the electric drift velocity $\bm{E}\times \bm{B}/B^2$ at a distance $r$ from the center of the star. A change in $y$ corresponds to a change in $\bm{E}$ which leads to a change in $\bm{E}\times \bm{B}/B^2$. The implication of opposite subpulse drifting suggests that the value of $y$ be generalized in terms of the polar, $\theta_b$, and azimuthal, $\phi_b$, angles in the magnetic frame \citep{MY16}. In this model, subpulse drifting is interpreted in terms of emission from overdense regions (subbeams), corresponding to $m$ emission areas, which flow relative to corotation under the influence of the $\bm{E}\times \bm{B}$ drift. Opposite subpulse drifting is then due to $\bm{E}$ being different in different emission states, corresponding to different values of $y$, across different parts of the profile resulting in the plasma flow rate to vary. The fact that observed drifting of subpulses can be related to the properties of the emission region allows the value of $y$ to be determined once the subpulse drift parameters are known. 

In our model, the drifting subpulses along a drift-band are interpreted as corresponding to a specific emission state described only by the parameter $y$ of a particular value. The variation in the emission state is implicitly assumed to be global in the magnetosphere, but we ignore it here. In addition, the magnetic field in the magnetosphere is treated as pure dipolar structure, and we focus on the pulsars with opposite subpulse drifting across pulse profile of small duty cycle (cf. Section \ref{sect:EmissionAreas}). Our investigation also neglect the mechanism that gives rise to opposite subpulse drifting. More information may be identified and included in the model once the triggering mechanism for the phenomenon is identified, but we do not do so here. We also allow variation in the number of subbeams ($m$), and we do not assume that subbeams on a carousel are evenly distributed.

The paper is organized as follows. We discuss the modeling and simulation for opposite subpulse drifting, together with the details of the three example pulsars, in Section \ref{sect:Model}. In Section \ref{sect:results}, we present the results obtained from our simulation. Discussion on the implications of our results is given in Section \ref{sect:discussion}, and we conclude the paper in Section \ref{sect:conclusions}. In the Appendices, we define the electromagnetic fields and the pulsar viewing geometry used in this paper.

\section{Modeling and simulation setup}
\label{sect:Model}

The subpulse drift rate is given by $P_2/P_3$, with $P_2$ signifies the horizontal separation between two subpulses within an individual pulse and $P_3$ characterizes the pulse period for the repeating pattern \citep{Pulsars1977}. In this section, we outline the model for drifting subpulses in a magnetosphere of multiple emission states as described by \citet{MY14a} and \citet{Yuen19b}. Most of the materials presented here on modeling are based on the two papers.

\subsection{Plasma flow}

The electric field for an obliquely rotating pulsar magnetosphere possessing multiple emission states has the form given by \citep{MY14a}
\begin{equation} \label{eq:EFieldNonCor}
\bm{E} = (1 - y\, \bm{b\,b}) \cdot \bm{E}_{\rm ind} + (1 - y)\, \bm{E}_{\rm pot},
\end{equation}
where $\bm{E}_{\rm pot} = -{\rm grad} \Phi_{\rm cor}$ represents the electric field in relation to the corotation charge density, and $y:=[0,1]$ represents a particular emission state. For $y=0$, $\bm{E} = \bm{E}_{\rm cor}$ is the corotation electric field given by equation (\ref{eq:Ecor}), and the emission state is described by the corotation model \citep{Goldreich1969}. The inductive electric field, $\bm{E}_{\rm ind}$, is due to an obliquely rotating magnetic dipole, which has the form given by equation (\ref{eq:Eind}). It contains a perpendicular and parallel components whose divergence has the form \citep{MY12}
\begin{equation}
{\rm div}\,\bm{E}_{\rm ind} = {\rm div}_\perp \bm{E}_{{\rm ind}\perp} + \frac{\partial{\it E}_{{\rm ind}\parallel}}{\partial s} = 0,
\end{equation}
where $s$ denotes distance along dipolar magnetic field lines. Here, $E_{{\rm ind}\parallel} = \bm{E}_{\rm ind} \cdot\bm{b}$ and $\bm{b}$ is the unit vector along the dipolar magnetic field lines. Taking into account the curvature of the field lines, we have \citep{MY12}
\begin{equation}
E_{\rm ind\parallel}=\frac{\mu_0}{4\pi}
\frac{\bm{b}\cdot[\bm{x}\times(\bomega_*\times\bm{\mu})]}{r^3},
\end{equation}
with
\begin{equation} \label{Amin3}
\bm{b}=
\frac{1}{\Theta}\bigg(2\cos\theta_m\hat{\bm{r}}-\frac{\partial\cos\theta_m}{\partial\theta}\hat{\btheta}
-\frac{1}{\sin\theta}\frac{\partial\cos\theta_m}{\partial\phi}{\hat\bphi}
\bigg)
\end{equation}
in spherical polar coordinates relative to the rotation axis and the unit vectors in radial, polar and azimuthal directions are represented by $\hat{\bm{r}},\hat{\btheta},\hat{\bphi}$, respectively. Here, $\Theta=(3\cos^2\theta_m +1)^{1/2}$, where $\theta_m=\cos\alpha\cos\theta + \sin\alpha\sin\theta \cos(\phi- \psi)$.

We assume a `minimal' state in the magnetosphere, designated by $y=1$, in which $\bm{E}_{{\rm ind}\parallel}$ is screened and $\bm{E}_{{\rm ind}\perp}$ has the same value as in the vacuum model. This requires a charge density be present to produce an electric field such that $\bm{E}_{\rm min} = -\bm{E}_{{\rm ind}\parallel}$. The divergence of $\bm{E}_{\rm min}$ implies a minimal charge density of the form \citep{MY14a}
\begin{equation} \label{eq:chargeDensityMin}
\rho_{\rm min}= -\varepsilon_0 {\rm div}\, (\bm{b}E_{{\rm ind}\parallel}).
\end{equation}
For states between the minimal state and the corotation state ($y=0$), a charge density is required to screen the electric field along the magnetic field lines in each state. For an oblique rotator the screening charge density, denoted by $\rho_{\rm sn}$, is given by \citep{MY14a}
\begin{equation} \label{eq:screeningChargeDensity}
\rho_{\rm sn} = y \rho_{\rm min} + (1-y) \rho_{\rm GJ},
\end{equation}
where 
\begin{equation} \label{eq:chargeDensityMin}
\rho_{\rm GJ}= -\varepsilon_0 {\rm div} \bm{E}_{\rm cor}
\end{equation}
is the corotation charge density \citep{Goldreich1969}. From Appendix \ref{sect:EMfields}, $\rho_{\rm sn}$ is proportional to $r^{-3}$ in dipolar field structure.

The plasma flow rate is the electric drift given by
\begin{equation} \label{eq:driftVel}
\bm{v}_{\rm dr} = \frac{\bm{E} \times \bm{B}}{ B^2} =  y \bm{v}_{\rm ind} + (1-y) \bm{v}_{\rm cor}.
\end{equation}
In the minimal state ($y=1$), $\bm{E} =\bm{E}_{{\rm ind}\perp}$ in the vacuum model and the electric drift velocity is given by $\bm{v}_{\rm ind} = \bm{E}_{{\rm ind}\perp} \times\bm{B}/B^2$. For $y=0$, the electric drift velocity in the region is given by $\bm{v}_{\rm dr}=\bm{v}_{\rm cor}=\bm{E}_{\rm cor}\times\bm{B}/B^2$ resulting in corotation in the region with the star. For a $y$ value between zero and unity, equation (\ref{eq:driftVel}) gives the electric drift velocity that combines a fraction y of the value due to the perpendicular component of the inductive electric field in the minimal state and that in the corotation model. Since $\bm{E}_{\rm ind}$, $\bm{E}_{\rm pot}$ and $\bm{B}$ are all functions of $\alpha$ (see Appendix \ref{sect:EMfields}), equation (\ref{eq:driftVel}) is also a function of $\alpha$. Furthermore, the allowance of different $y$ values in an emission region implies that different emission states can coexist in the emission region. Dividing equation (\ref{eq:driftVel}) by $r$ gives the angular velocity, $\bm{\omega}_{\rm dr}$, in the form  
\begin{equation} \label{eq:driftAngVel}
\bm{\omega}_{\rm dr} = y \bm{\omega}_{\rm ind} + (1-y) \bm{\omega}_{\rm cor}
\end{equation} 
for the plasma flow in the magnetosphere as a function of $y$. For a magnetosphere in oblique rotation, the allowed emission states ($y$) are of the form defined by equation (\ref{eq:driftAngVel}).

\subsection{Emission areas}
\label{sect:EmissionAreas}

\begin{figure}
\centering  
\includegraphics[width=1\columnwidth]{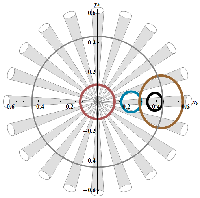}
\caption{Plot showing different trajectories of the visible point in the magnetic frame where the magnetic pole is at the origin. The four trajectories are constructed using $\{\zeta,\alpha\}=\{6^\circ,20^\circ\}$ (blue), $\{5^\circ,35^\circ\}$ (black), $\{15^\circ,40^\circ\}$ (brown), and $\{10^\circ,0^\circ\}$ (red). Different trajectories have different sizes and shapes, and their centers can locate away from the magnetic pole and not including the magnetic pole as shown by the first three trajectories. The case where the trajectory centers at, and encloses, the magnetic pole is shown by the red trajectory. The circle in gray represents the boundary of an open-field region at $r=0.2 r_L$. All trajectories are either partly (black and brown) or entirely (blue and red) enclosed in the region. Also shown is the structure of radial spokes depicted in gray circular cones emanating from the origin.} 
\label{fig-trajectories_spokes}
\end{figure}

We consider emission from subpulses as coming from discrete areas of overdense plasma whose formation is due to the existence of a standing wave at a specific spherical harmonic caused by an instability in the magnetosphere \citep{CR04, GMM+05}. This forms a pattern of anti-nodes (overdense plasma) and nodes, which varies in proportional to $\cos(m\phi_b)$ around the magnetic axis. We assume that observable radio emission originates from $m$ emission areas restricted to the anti-nodes. One arrangement of the emission areas involves their locations be spaced equally in azimuth around the magnetic axis \citep{GS00}. This implies that the arrangement is independent of the polar angle, $\theta_b$, in the magnetic frame (designated by the subscript $b$), and locally independent of height, $r$. Therefore the emission areas align along the radial direction when projected onto a surface of constant $r$ resulting in a structure of radial spokes as shown in Figure \ref{fig-trajectories_spokes}. 

The emission is assumed to occur in a geometry in which radiation at the source point is directed along the local dipolar magnetic field line \citep{Hibschman2001} and radiation arises only within the open-field region \citep{Cordes78, KG03}. Description of the geometry is given in Appendix \ref{sect:PulsarVisibility}. There are two angles that define any given pulsar in this geometry, which are the $\zeta$ and $\alpha$. An explicit solution for the geometry is defined by equation (\ref{eq:EmissionPtPhiB}), which gives the angular location for the point of visible emission in terms of the polar, $\theta_{b{\rm V}}$, and azimuthal, $\phi_{b{\rm V}}$, angles in the magnetic frame as a function of $\psi$ for given $\zeta$ and $\alpha$. The visible point moves at an angular speed, $\omega_{\rm V}$, that varies as the pulsar rotates with the lowest speed dependent on $\alpha$. The visible point then traces a closed curve after one pulsar rotation, referred to here as the trajectory of the visible point (see Figure \ref{fig-trajectories_spokes}), or simply ``trajectory'' where there is no confusion. In the observer's frame, the trajectory is expressed in terms of the polar, $\theta_{\rm V}$, and azimuthal, $\phi_{\rm V}$, angles relative to the rotation axis. In general, the trajectory is not circular or centered at the magnetic axis. We designate the emission areas that are cut by the trajectory of the visible point as emission spots. Observable emission then requires the emission spots be coincided with the trajectory of the visible point that must lie inside the open-field region. It implies that the number of emission spots is estimated along the trajectory of the visible point. As different trajectories (with different combination of $\zeta,\alpha$) cut the emission areas at different $\theta_{\rm V},\phi_{\rm V}$, the visible emission across a profile window is also dependent on $\theta_{\rm V}$ and $\phi_{\rm V}$ or $\theta_{b{\rm V}}$ and $\phi_{b{\rm V}}$. This also implies that the perceived arrangement of emission spots is unique for different pulsars. In Figure \ref{fig-trajectories_spokes} for example, a special arrangement is for $\alpha=0$, in which the red trajectory cuts all the spokes equally at identical $\theta_{b{\rm V}}$ around the magnetic axis. In this case, the estimation of emission spots along the trajectory gives the actual number of spokes around the magnetic axis regardless of the width of the profile window. For oblique rotators, the estimation of emission spots located at similar $\theta_{b{\rm V}}$ is realized only for narrow pulse profiles where variation of $\theta_{b{\rm V}}$ is small along the trajectory within the profile window.

The apparent value of $\omega_{\rm dr}$ is determined by the components projected onto the trajectory of the visible point. For oblique rotation, $\omega_{\rm dr}$ varies as a function of $\psi$, and it changes as $y$ changes for a given $\psi$.

\subsection{Subpulse drifting and emission state}

The assumption of the emission area being fixed to the magnetospheric plasma implies that the emission area has an angular velocity identical to that of the plasma given by equation (\ref{eq:driftAngVel}). The determination of $\omega_{\rm dr}$ at locations defined by the trajectory of the visible point implies that $P_2$ represents the time between the visible point coinciding with neighboring anti-nodes (emission spots). Consider the case when the motion of the visible point is ignored, $m$ emission spots will pass through the line of sight in the time given by $2\pi/\omega_{\rm dr}$ that the plasma takes for one complete rotation. The separation of consecutive emission spots would then be observed with an interval given by $2\pi/m\omega_{\rm dr}$. When taking $\omega_{\rm V}$ into account, the interval is modified into \citep{Yuen19b}
\begin{equation}\label{eq:P2}
P_2 (y)= \frac{2\pi}{m\omega_{\rm dr}-\omega_{\rm V}},
\end{equation}
with $\omega_\star P_2$ in radians. The anti-nodes and the visible point are rotating at different angular velocities and both are different from $\omega_\star$. The traditional picture of a plasma-filled magnetosphere \citep{Goldreich1969, RudermanSutherland1975} suggests that the delay between identical repeating patterns ($P_3$) is dependent on $\omega_{\rm dr} -\omega_{\rm cor}$. A stationary pattern unchanged in sequence of pulses is signified by $\omega_{\rm dr}=\omega_{\rm cor}$ (at $y=0$), and a difference between $\omega_{\rm dr}$ and $\omega_{\rm cor}$ will result in a slowly changing pattern in consecutive pulses and repeats after several pulsar rotations. Since the same momentary drift pattern is seen by the observer, this suggests that the motion of the visible point is irrelevant, thus giving \citep{Yuen19b}
\begin{equation}\label{eq:P3}
P_3 (y)= \frac{2\pi}{m(\omega_{\rm dr} -\omega_{\rm cor})},
\end{equation}
with both $P_2$ and $P_3$ determined on the trajectory of the visible point. The subpulse drift rate is given by 
\begin{equation}\label{eq:driftrateTime}
\frac{P_2}{P_3} (y)= \frac{m(\omega_{\rm dr} -\omega_{\rm cor})}{m\omega_{\rm dr}-\omega_{\rm V}}.
\end{equation}
From equations (\ref{eq:P2})--(\ref{eq:driftrateTime}), the values of $P_2$, $P_3$ and the associated drift rate will change with a change in $m$ or $y$ for a pulsar. 

Traditional convention assumes that $P_3$ is always positive, and the subpulse drift direction is signified by the sign in $P_2$ \citep{Wetal06}. For subpulses appearing progressively later in successive pulses, the tracks traced by the drifting subpulses tilt forward resulting in a positive slope and $P_2$ is defined as positive giving a positive drift, and vice versa for negative drift. In our model, a change in the $y$ value corresponds to a change in $\omega_{\rm dr}$, which leads to variation in the sign of ($\omega_{\rm dr} -\omega_{\rm cor}$). From equation (\ref{eq:P3}), $\omega_{\rm dr} >\omega_{\rm cor}$ indicates a positive $P_3$ and the plasma flow is ahead of corotation. The change in the longitudinal phase of an emission spot between two consecutive pulses is positive, and the emission spot will appear moving towards later longitudinal phases in successive pulses. The subpulses trace a forward-tilting track with a positive slope giving a positive drift-rate, where $P_2$ is signified in the traditional convention by a value in positive. On the contrary, a negative $P_3$ means $\omega_{\rm dr} <\omega_{\rm cor}$ such that the plasma flow lags the corotation. This results in a negative change in the longitudinal phase of an emission spot between two consecutive pulses, and the emission spot will appear moving towards earlier longitudinal phases in successive pulses. This gives a negative slope for the drift tracks and a negative drift-rate, where $P_2$ is assumed negative in the traditional convention. The prediction of drifting subpulses in opposite directions would then require the sign of ($\omega_{\rm dr} -\omega_{\rm cor}$) in equation (\ref{eq:P3}) be different across different longitudinal phases such that the flow of plasma reverses relative to corotation.

\subsection{Simulation setup}
\label{sect:Setup}

We apply the model to PSRs B0525+21, B1929+10 and B0052+51 for investigation of the emission properties in the emission regions as revealed by the opposite subpulse drifting. For each of the three pulsars, drifting subpulses are observed in both components of the integrated pulse profile with the drift direction opposite to each other \citep{Wetal06}. The values of $\zeta$ and $\alpha$ are known only for the first two pulsars from \citet{LyneManchester1988}. Details of the pulsars and the parameters for the opposite subpulse drifting are reproduced in blocks II and III in Table \ref{table:driftStats} for easy references. 

Since drifting subpulses are found in two different parts of the profile in all three pulsars, we divide each profile into two component regions, A and B, of equal width. Simulation is then performed using equations (\ref{eq:P2})--(\ref{eq:driftrateTime}) to search for the values of $m$ and $y$ that give the $P_2$, $P_3$ and the drift rate within the respective observed uncertainty for each pulsar based on $m := [1,45]$, in step of 1, and $y:=[0,1]$, in step of $10^{-4}$. The calculation involves determining $\omega_{\rm dr}$, $\omega_{\rm cor}$ and $\omega_{\rm V}$ along the trajectory of the visible point over the respective profile width with division of $0.1^\circ$. For the pulsar with unknown $\zeta$ and $\alpha$, the simulation also includes an extra consideration of possible ranges for $\zeta$ and $\alpha$ that give the required drift parameters. To determine $\alpha$, search is performed from $1^\circ$ to $85^\circ$, in step of $1^\circ$, with an impact parameter $\beta = \zeta-\alpha \leq |5^\circ|$ under the maximum height of $0.2 r_{\rm L}$. Once a calculation produces simulated values that match the observed values within the corresponding uncertainties, the simulated values are then weighted based on the observed drift parameter with the least reported uncertainty. For the three pulsars, it is in the $P_3$ value \citep{Wetal06}. For example, a simulated $P_3$ value that falls within the uncertainty of the observed value receives a weight proportional to $1 - |x|$, where $x$ is the interval between the observed and simulated values, normalized by the maximum reported uncertainty. Therefore, a result that matches the observed value more closely will receive more weight, and each simulated value from the same calculation will receive the same weight. At the end of the whole simulation, a weighted average is determined from each of the simulated values. Once the value of $y$ is decided for a component region, equation (\ref{eq:screeningChargeDensity}) is used to obtain the charge density ($\rho_{\rm sn}$) for that region.

\section{The results}
\label{sect:results}

The results of our simulation are shown in blocks IV and V in Table \ref{table:driftStats}. In interpreting our results, we assume that emission detected at the same observing frequency comes from emission spots that are located on the same carousel \citep{SMK05}.

\begin{table*}[ht]
\huge
\centering
\begin{adjustbox}{width=17.5cm,right}
\setlength\extrarowheight{17.pt}
\begin{tabular}{l|cccc|ccc|ccc|ccc}
	\hline
	
\multicolumn{1}{l|}{I. Name} & \multicolumn{4}{c|}{II. Geometry} & \multicolumn{3}{c|}{III. Observation} & \multicolumn{3}{c|}{IV. Simulation} & \multicolumn{3}{c}{V. Emission properties} \\
\hline

${\rm PSR}$ & $\beta^\circ$ & $\alpha^\circ$ & $\Delta\psi^\circ$ & {\rm Reg.} & $P_2(^\circ)$ & $P_3\, (P_1)$ & {\rm Drift rate} & $P_2(^\circ)$ & $P_3(P_1)$ & {\rm Drift-rate} & $m$ & $y$ & $\rho_{\rm sn}/\rho_{\rm GJ}$ \\
\hline

\multirow{2}{*}{\rm B0525+21} & \multirow{2}{*}{0.7} & \multirow{2}{*}{23.2} &\multirow{2}{*}{20} & {\rm A} & $-20^{+2}_{-9}$ & $3.8\pm 0.7$ & $-1.4^{+0.3}_{-0.7}$ & $-23\pm 4$ & $3.5\pm 0.2$ & $-1.6\pm 0.3$ & $28\pm 2$ & $0.139\pm 0.011$ & $1.025\pm 0.011$ \\
&&&& {\rm B} & $50^{+55}_{-10}$ & $3.7\pm 0.4$ & $3.6^{+4.0}_{-0.8}$ & $73\pm 19$ & $3.6\pm 0.1$ & $5.2\pm 1.4$ & $28\pm 2$ & $0.332\pm 0.047$ & $1.129\pm 0.047$ \\
\hline

\multirow{2}{*}{\rm B1929+10} & \multirow{2}{*}{4.0} & \multirow{2}{*}{6.0} &\multirow{2}{*}{20} & {\rm A} & $90^{+140}_{-8}$ & $9.8\pm 0.8$ & $40.5^{+63.1}_{-4.9}$ & $123\pm 37$ & $9.8\pm 0.1$ & $55.4\pm 16.8$ & $28\pm 2$ & $0.574\pm 0.061$ & $0.459\pm 0.061$ \\
&&&& {\rm B} & $-160^{+10}_{-100}$ & $4.4\pm 0.1$ & $-160.5^{+10.7}_{-100.4}$ & $-180\pm 27$ & $4.4\pm 0.1$ & $-180\pm 27$ & $32\pm 2$ & $0.675\pm 0.038$ & $0.367\pm 0.038$ \\
\hline
\hline

\multirow{2}{*}{\rm B0052+51} & \multirow{2}{*}{$\leq|5|$} & \multirow{2}{*}{$11\pm 1$} & \multirow{2}{*}{15} & {\rm A} & $-75^{+50}_{-50}$ & $4\pm 2$ & $-8.9^{+7.4}_{-7.4}$ & $-50\pm 16$ & $4.0\pm 1.1$ & $-5.8\pm 1.0$ & $5\pm 1$ & $0.847\pm 0.067$ & $0.153\pm 0.067$ \\
&&&& {\rm B} & $30^{+70}_{-7}$ & $5\pm 1$ & $2.8^{+6.6}_{-0.9}$ & $23\pm 13$ & $4.8\pm 0.6$ & $2.2\pm 1.3$ & $15\pm 1$ & $0.313\pm 0.075$ & $0.687\pm 0.075$ \\
\hline
\end{tabular}
\end{adjustbox}
\caption{Details of the opposite subpulse drifting for the three pulsars used in this paper are shown in blocks II and III, and the results of our simulation are shown in blocks IV and V. In block II, the values of $\beta$ and $\alpha$, for PSRs 0525+21 and B1929+10 \citep{LyneManchester1988} and for PSR B0052+51 (simulated), are given in columns 2 and 3, respectively. The pulse-width, measured at about 10\% of the maximum intensity, and the different component regions for each pulsar are listed in the fourth and fifth columns, respectively. The observed drift parameters are obtained from \citet{Wetal06} and reproduced in block III, with the drift rate expressed in deg\,s$^{-1}$. The simulated $P_2$, $P_3$ and the drift-rate for each component region are given in block IV, and the corresponding weighted values for $m$, $y$, and the associated charge density, are given in block V. Note that the values of $m$ are rounded up to the nearest integer, and $\rho_{\rm sn}$ is expressed in units of the Goldreich-Julian value.}
\label{table:driftStats}
\end{table*}

\subsection{Simulated drift parameters}

Block IV shows the values for $P_2$, $P_3$ and the drift rate obtained from our simulation for the drifting subpulses in each component region in the three pulsars. The change in the drift rate across a profile is due to difference in the emission state ($y$) and, in some cases, the subpulse number ($m$) in different component regions of a profile (see Section \ref{sect:emissionProperties}). The prediction of $\alpha$ for PSR B0051+52 is $11^\circ\pm 1^\circ$ from our simulation.

\begin{figure}
\centering  
\includegraphics[width=1\columnwidth]{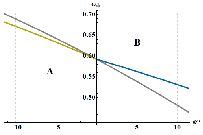}
\caption{Plot showing variations in the plasma flow rate between the two component regions relative to corotation across the profile of PSR B0525+21. The calculation uses $\zeta=23.9^\circ,\alpha=23.2^\circ$ and also based on the simulated $y$ and $m$ values (omitting the uncertainties) in each region given in Table \ref{table:driftStats}. The plasma flow rate is normalized to corotation, which are indicated by the curves in yellow and blue in regions A and B, respectively, and corotation is represented by the curve in gray. The sign of averaged $P_3$ changes from negative to positive in regions from A to B. The boundaries of the pulse profile are indicated by the two vertical dashed lines.} 
\label{fig-driftrate_0525}
\end{figure}

\begin{figure}
\centering  
\includegraphics[width=1\columnwidth]{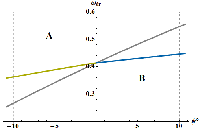}
\caption{Similar to Figure \ref{fig-driftrate_0525}, this plot shows the normalized plasma flow rate relative to corotation in each of the two component regions for PSR B1929+10 using $\zeta=10^\circ, \alpha=6^\circ$.} 
\label{fig-driftrate_1929}
\end{figure}

\begin{figure}
\centering  
\includegraphics[width=1\columnwidth]{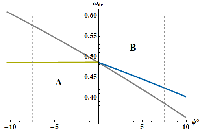}
\caption{Similar to Figure \ref{fig-driftrate_0525} but for PSR B0052+51 using $\alpha=11^\circ$ and assuming $\zeta=15^\circ$.} 
\label{fig-driftrate_0052}
\end{figure}

In our model, subpulses drift when $\omega_{\rm dr} \neq\omega_{\rm cor}$ and the drift would appear in opposite direction when ($\omega_{\rm dr} -\omega_{\rm cor}$) changes sign in different component regions of a profile. Figures \ref{fig-driftrate_0525}--\ref{fig-driftrate_0052} show the plasma flow rate normalized to corotation and plotted coinciding with the observed longitudinal phase of the profile using the weighted values of $m$ and $y$ for each of the component regions as indicated in Table \ref{table:driftStats}. For PSR B0525+21, the corotation is higher than the plasma flow rate in region A, and reverses in region B, with the average values of $(\omega_{\rm dr}/\omega_{\rm cor})$ given by 0.98 and 1.05, respectively. This indicates that the drift direction is opposite between the two regions in the way that it is negative in the former region and reverses in the latter. This is similar for PSR B0052+51, whose $(\omega_{\rm dr}/\omega_{\rm cor})=$ 0.93 and 1.05 for regions A and B, respectively. The different drift directions in PSR B1929+10 corresponds to different $(\omega_{\rm dr}/\omega_{\rm cor})$ in regions A and B, which has the average values of 1.14 and 0.89, respectively. This implies a positive drift in region A but reverses in region B. In Table \ref{table:driftStats}, we conform with the traditional convention by indicating the drift rate direction in $P_2$.

\subsection{The different emission properties}
\label{sect:emissionProperties}

The values of $m$ and $y$ obtained from our simulation based on the drifting subpulses in each of the component regions are shown in block V. For the three pulsars in our sample, the $y$ values are different in different component regions. This indicates that opposite subpulse drifting in the three pulsars originates from different emission properties with each corresponds to a particular emission state that occupies a specific part of the emission region. The average values of $y$ are $0.24\pm 0.04$, $0.62\pm 0.07$ and $0.58\pm 0.15$ for PSRs B0525+21, B1929+10 and B0052+51, respectively. Since $y \neq 0$, the plasma flow is not in corotation for all three pulsars.

The number of emission spots on the carousel for a pulsar, as revealed by the value of $m$, may or may not vary across different component regions of a pulsar. For PSRs B1929+10 and B0052+51, the $m$ shows a preference for a different value in different component regions. However, the two component regions in PSR B0525+21 share a similar $m$ value within the uncertainty. From top to bottom, the average values of $m$ predicted from the different component regions for each pulsar are $28\pm 3$, $30\pm 3$ and $10\pm 3$, respectively. This indicates that $m$ varies with pulsars, which is consistent with some investigations that predict different subpulse number for different pulsars \citep{GMG03, ELG+05, SMS+07,MR08}. In addition, the variation of $m$ does not show clear correlation with $\alpha$. From the prediction of similar $m$ values shared between regions A and B regardless of $y$ in PSR B0525+21, we also conclude that the value of $m$ does not show correlation with $y$. This implies that the number of subpulses on a carousel is not related to the emission state in the emission region. The average $m$ values from components with positive and negative drift rates are identical within the uncertainty at $m=24\pm 3$ and $m=22\pm 5$, respectively. The overall average (both regions included) value of $m$ for the three pulsars is $23\pm 6$, which is consistent with $20$ predicted by \citet{MR08}.

It is straightforward to determine the overall charge density, $\rho_{\rm sn}$, for each component region using equation (\ref{eq:screeningChargeDensity}) once the value of $y$ is known for that region. The results are expressed in units of $\rho_{\rm GJ}$ and shown in the last column in block V. The value of $\rho_{\rm sn}/\rho_{\rm GJ}$ is not unity indicating that the charge density deviates from the Goldreich-Julian value in all component regions. For the three pulsars, the differences in $\rho_{\rm sn}/\rho_{\rm GJ}$ between different component regions suggest that the charge density is not evenly distributed across the observable emission region. In addition, the charge density in an emission region can be higher, as in PSR B0525+21, or lower than the Goldreich-Julian value. From the three pulsars, the distribution of charge density in an emission region does not show clear correlation with $\alpha$. 

\section{Discussion}
\label{sect:discussion}

In this section, we explore the implications of our results on subpulse distribution and different emission states in the emission region as revealed by opposite subpulse drifting. 

\subsection{Distribution of subpulses}
\label{sect:subpulseDistribution}

A question in the traditional models for polar cap emission concerns the arrangement of subpulses and their relation to $r$. Both uniform and non-uniform distributions of subpulses have been proposed \citep{Manchester1995, GS00}. An implication of different $m$ between different profile component regions described in Section \ref{sect:emissionProperties} suggests that the distribution of subpulses, or the subpulse density, varies across the observable emission region. We estimate the ratio of the subpulse density as follows. First, the assumption of pulsar radio emission originated from outflow of relativistic pair plasma along open magnetic field lines in two-stream instability would imply that radio luminosity varies in proportional to the plasma density \citep{Lyubarskii96}. In our model, the plasma density is given by $\lambda\rho_{\rm sn} = n_+ + n_-$ \citep{MY12}, where $\lambda$ and $n_\pm$ are the pair multiplicity and the electrons and positrons, respectively. Here, $\rho_{\rm sn}$ represents the charge density associated with a particular $y$ value in the region. Assuming identical $\lambda$ for all $y$ would imply that the observed profile intensity is correlated with $\rho_{\rm sn}$ such that $I \propto\rho_{\rm sn}$. Furthermore, the dependence of both $\rho_{\rm min}$ and $\rho_{\rm GJ}$ on $r$ makes it possible to estimate the ratio of the emission heights between any two component regions, say A and B, denoted by $r_A/r_B$, using equation (\ref{eq:screeningChargeDensity}) for known ratio of the observed peak profile intensity, $I_A/I_B$. For narrow profile, the angle from the magnetic axis to the different emission spots cut by the trajectory of the visible point ($\theta_{b{\rm V}}$) is approximately identical across the profile window regardless of the variation in height. For the three pulsars in our sample, the average change in $\theta_{b{\rm V}}$ across the profiles is $2.3^\circ$. Then, the ratio of the subpulse density on the circular carousel between the two regions as predicted by the different drifting subpulses can be estimated by $D_A/D_B = (m_A/m_B) (r_B/r_A)$ using the corresponding $m$ values. A uniform distribution of subpulses is signified by $D_A/D_B =1$, whereas $D_A/D_B\neq 1$ indicates that the subpulse density is different between the two component regions and the subpulse distribution varies across the observable emission region. Thus $D_A/D_B$ serves as a measure for the uniformity of the subpulse distribution across the observable region represented by the two component regions. The results are shown in Table \ref{table:driftgeo}.\\

\noindent{\it PSR B0525+21:} 

\noindent Our results show that this pulsar displays preference for higher profile intensity to emission that comes from region B, where $\rho_{\rm sn}$ is higher. This agrees with $I \propto\rho_{\rm sn}$ suggesting that the plasma density is higher in region B. In addition, the value of $D_A/D_B$ is close to unity indicating that constant subpulse density is preferred across the two regions. For $r_A/r_B\sim 1$ within the uncertainty, this implies that the arrangement of subpulses on the carousel is likely of a uniform manner in this pulsar. \\

\noindent{\it PSR B1929+10:}

\noindent The emission from the two regions of this pulsar show preference for coming from different heights, with $r_A>r_B$, but each with similar $\rho_{\rm sn}$ within the uncertainty. In addition, the higher values of $m$ and $y$ in region B are associated with higher profile intensity. This gives $D_A/D_B\neq 1$ with region B displaying higher subpulse density. This implies a non-uniform distribution of subpulses on the carousel.\\

\noindent{\it PSR B0052+51:} 

\noindent The emission from the two component regions originates from different heights with region A being lower and the profile intensity is also lower. In addition, the lower intensity corresponds to a lower value in both $m$ and $\rho_{\rm sn}$ than that in region B. Furthermore, $D_A/D_B < 1$ implying that the predicted subpulse density is greater in region B, which is consistent with $m$ displaying higher values in the region. The prediction of $D_A/D_B\neq 1$ suggests that the distribution of subpulses across the observable region is also different. This shows that the arrangement of subpulses on the carousel is of a non-uniform manner in this pulsar.\\

In summary, distribution of subpulses on a carousel can be either uniform or non-uniform as revealed by opposite subpulse drifting in our sample.

\begin{table}
\small
\centering
\setlength\extrarowheight{4.1pt}
\setlength{\tabcolsep}{5.2pt}
\begin{tabular}{l||ccc|c}
	\hline
${\rm PSR}$ & $I_A/I_B$ & $r_A/r_B$ & $D_A/D_B$ & {\rm Type} \\
\hline
\hline

{\rm B0525+21} & 0.8 & $1.04\pm 0.17$ & $0.95\pm 0.18$ & {\rm U} \\
\hline

\rm B1929+10 & 0.8 & $1.28\pm 0.15$ & $0.68\pm 0.10$ & N \\
\hline

\rm B0052+51 & 0.6 & $0.72\pm 0.18$ & $0.46\pm 0.15$ & N \\
\hline

\hline
\end{tabular}
\caption{Information related to subpulse distribution in the emission region derived from block V in Table \ref{table:driftStats} for the three pulsars. The last column indicates the type of subpulse distribution as suggested by our results, with U=Uniform and N=Non-uniform.}
\label{table:driftgeo}
\end{table}

\subsection{Distribution of emission states}
\label{sect:DistributionEmissionStates}

In the \citet{RudermanSutherland1975} model, the existence of a potential difference in the vacuum gap near the polar cap implies that plasma density deviates from $\rho_{\rm GJ}$ and the plasma flow rate changes from corotation across the gap \citep{MY16}. In our model, this suggests that $y$ changes as a function of $r$, or $\theta_b$ with $\theta_b = \sin^{-1}\sqrt{r/r_0}$ on a dipolar field line, where $r_0$ is the field-line constant, implying $y=y(\theta_b)$. This is seen in PSRs B1929+10 and B0052+51 in the way that $y$ increases as $r$ decreases, as shown in Tables \ref{table:driftStats} and \ref{table:driftgeo}. The variation shows a preference for more deviation from corotation at lower height within the observable emission region. For the two pulsars, emission from different component regions corresponds to different emission states and each locates at a different height. In addition, the association of unique drift rate and direction with particular $y$ value at different parts of a profile indicates that $y$ also varies as a function of the longitudinal phase, $\phi_b$. This implies $y=y(\theta_b,\phi_b)$. Another distribution is shown in PSR B0525+21 where different emission states can coexist at similar height but at different longitudinal phases, that is $\phi_b$ varies while $\theta_b$ is a constant. Since similar emission states can exist in pulsars with different $\alpha$, as for region B in PSRs B0525+21 and B0052+51, it suggests that the distribution of emission state in the emission region of pulsars with opposite subpulse drifting is not likely correlated with $\alpha$.

\subsection{Changes in emission state}

Variations in emission state are also found in another category of drifting subpulses. Similar to the three pulsars in this paper, the drift patterns in this category are unique to different emission states and changes occur via emission state switching. A well-known example is PSR B0031$-$07, whose drifting subpulses demonstrate three different modes of drift rates with switching between different drift modes \citep{SMK05}. However, the different drift patterns do not function at different ranges of longitudinal phase but at different times. For the pulsar, the observable emission region can also accommodate different emission states but the manifestation of each is exclusive from the other meaning that only one drift pattern is observable at any one time. This implies that different emission states cannot coexist. However, the cyclical change between the different modes of drift rates indicates that the pulsar behaves as if it ``remembers'' the different emission states. In our model, this implies $y=y(t)$ for the pulsar, and emission state switching affects the whole observable emission region. This suggests $y=y(\theta_b, \phi_b, t)$ in general, which implies that the two categories of drifting subpulses may be manifestations of the same underlying mechanism. Furthermore, if $t$ is also a variable in opposite subpulse drifting pulsars, then the different drifting subpulses may also exhibit switching which will demonstrate as sudden changes in the drift rate. The rarity of pulsars with subpulse drift-mode switching and opposite subpulse drifting may indicate that such phenomena are not common in radio pulsars. Since our results suggest that the value of $y$ is not related to $\alpha$, and if $\alpha$ is assumed correlating with the evolution of the star \citep{BGI84}, it would suggest that the phenomena can exist in pulsars of different evolutionary stages thus increasing the chance of their discovery. It may be that opposite subpulse drifting and drift-mode switching is a small effect in some pulsars thereby requiring high sensitivity equipments for the detection. With the operation of the 500-m Aperture Spherical radio Telescope (FAST) \citep{LWQ+18} and the future Square Kilometer Array (SKA), there is little doubt that higher quality data will be available which can reveal drifting subpulses of increasingly complicated details thus deepening our understanding of pulsar radio emission mechanism.

\section{Conclusions}
\label{sect:conclusions}

We have investigated properties of the emission region as revealed by opposite subpulse drifting based on the rotating carousel model incorporating the model for pulsar magnetosphere of multiple emission states. A quantitative definition of an emission state is through the plasma flow, $\omega_{\rm dr}$, in the magnetosphere, which is parameterized by $0\le y\le1$ with the inferred emission state corresponds to a particular value in $y$. The observable drift of subpulses in an emission state is due to $\omega_{\rm dr}$, as induced by the $\bm{E}\times\bm{B}$ drift, that deviates from corotation. A change in the emission states ($y$) corresponds to a change in $\bm{E}$ causing $\omega_{\rm dr}$ to vary resulting in the diversity of drifting subpulses across a profile. We interpret the subpulses as emission from $m$ discrete areas that are arranged in azimuth around the magnetic axis. Our simulation of opposite subpulse drifting using a magnetosphere of pure dipolar magnetic field structure is possible only if the assumption for the existence of multiple emission states in the emission region is allowed. This is because changes in the emission state, as revealed by variation in the value of $y$ in our model, result in changes in the emission properties and hence the subpulse drift characteristics. The model is applied to three pulsars. Our results show that each of the different subpulse drift rates and the corresponding drift direction at a particular range of longitudinal phase can be associated with a particular value of $y$ in the emission region. This suggests that different emission states can coexist in the emission region and they operate concurrently. Furthermore, the distribution of different emission states is functions of azimuthal and polar angles around the magnetic axis, but not function of the obliquity angle. In addition, the association of different emission properties with different emission states makes it possible to discern the different emission environments, such as the estimation of the charge density, in relation to the phenomenon across the observable emission region. The model also allows the identification of the subpulse number ($m$) as revealed by the different drifting subpulses in different component regions. Together with the ratio of the peak profile intensity between the two regions of different drifting subpulses, we also estimate the ratio of subpulse density between different component regions from which the distribution of subpulses on a carousel is inferred. For pulsars with opposite subpulse drifting, we find that the number of subpulses on a carousel is not related to the obliquity angle or the emission properties of a particular emission state. In addition, the distribution of subpulses on a carousel is pulsar dependent, which can be uniform or non-uniform.

\acknowledgments
We thank Wenming Yan, Zhigang Wen and Willem Baan for useful discussion. We also thank the referee for useful comments. RY is supported by the 2018 Project of Xinjiang Uygur Autonomous Region of China for Flexibly Fetching in Upscale Talents, and Natural Science Foundation of China (grant No. U1838109, 11873080, 12041301), and partly supported by Xiaofeng Yang's Xinjiang Tianchi Bairen project and CAS Pioneer Hundred Talents Program.

\bibliographystyle{spr-mp-nameyear-cnd}

\appendix

\section{The electromagnetic fields}
\label{sect:EMfields}

The electric and magnetic fields for a rotating magnetic dipole in vacuo are \citep{MY14a}
\begin{equation}\label{eq:Eind}
\bm{E}_{\rm ind}=\frac{\mu_0}{4\pi}\left[\frac{\bm{x}\times \dot{\bm{\mu}}}{r^3}+\frac{\bm{x}\times \ddot{\bm{\mu}}}{r^2c}\right],
\end{equation}
and
\begin{equation} \label{eq:RotationalMagneticField}
\bm{B} = \frac{\mu_0}{4\pi} \Bigg[ \frac{3\bm{xx}\cdot\bm{\mu} - r^2\bm{\mu}}{r^5} + \frac{3\bm{xx}\cdot \dot{\bm{\mu}} - r^2 \dot{\bm{\mu}}}{r^4c} + \frac{\bm{x} \times ( \bm{x} \times \ddot{\bm{\mu}} )}{r^3 c^2} \Bigg],
\end{equation}
where $c$ is the light speed and $\mu_0$ is the vacuum permeability. Here, $\bm{x}$ and $r$ are the position vector and radial distance from the stellar center, respectively, and the time-dependent magnetic dipole is represented by $\bm{\mu}$. The dipolar term in the magnetic field is given by $1/r^3$ term, and the radiative terms are denoted by the terms $\propto 1/r^2$ and $\propto 1/r$. In spherical coordinates, the expression for the first term in $\bm{E}_{\rm ind}$ has the form given by
\begin{equation} \label{EindFor}
\left(
\begin{array}{c}
E_{{\rm ind},r}\\
E_{{\rm ind},\theta}\\
E_{{\rm ind},\phi}
\end{array}
\right)={\frac{\mu_0 \mu\omega \sin\alpha}{4\pi r^2}}
\left(
\begin{array}{c}
0\\
-\cos(\phi-\psi)\\
\cos\theta\sin(\phi-\psi)
\end{array}
\right),
\end{equation}
which is nonzero for an oblique rotator ($\alpha\neq 0$). Both $\bm{E}_{\rm ind}$ and $\bm{B}$ can be determined uniquely at a location of a given coordinates for known $\zeta$ and $\alpha$ (see Appendix \ref{sect:PulsarVisibility}). 

In a plasma-filled magnetosphere with negligible particle inertia and infinite conductivity, the electric field vanishes in the co-moving frame of the plasma giving the corotation electric field as
\begin{equation}\label{eq:Ecor}
\bm{E}_{\rm cor} = -(\bomega_\star \times \bm{x}) \times \bm{B},
\end{equation}
where $\bomega_\star$ is the angular velocity of the star. For an obliquely rotating magnetosphere, equation (\ref{eq:Ecor}) can be written in the form \citep{Hones1965, Melrose1967}
\begin{equation} \label{eq:Ecor_obli}
\bm{E}_{\rm cor} = -{\rm grad}\,\Phi_{\rm cor} - \frac{\partial\bm{V}}{\partial t},
\end{equation}
where $\bm{V}$ representing the vector potential of a rotating magnetic dipole \citep{MY14a}, and $\bm{E}_{\rm ind} = -\partial\bm{V}/\partial t$. In dipolar field structure, $\bm{E}_{\rm cor}$ is perpendicular to the field lines, and has a component only along the radial and polar directions in spherical coordinates. 

\section{The emission geometry}
\label{sect:PulsarVisibility}

In this Appendix, we will summarize the viewing geometry used in this paper \citep{YM14}. We assume an arrangement of the rotation and magnetic axes of a pulsar in the way that $\hat\bomega_\star ={\hat{\bf z}}$ and $\hat{\bf m} ={\hat{\bf z}}_b$, respectively, in Cartesian coordinates of unit vectors given by ${\hat{\bf x}},{\hat{\bf y}},{\hat{\bf z}}$ and ${\hat{\bf x}}_b,{\hat{\bf y}}_b ,{\hat{\bf z}}_b$. The transformation between the unit vectors is given by
\begin{equation}
\left(
\begin{array}{c}
\hat{\bf x}_b\\{\hat{\bf y}}_b\\{\hat{\bf z}}_b
\end{array}
\right)=
\textbf{\textsf{R}}
\left(
\begin{array}{c}
{\hat{\bf x}}\\{\hat{\bf y}}\\{\hat{\bf z}}
\end{array}
\right)
\hspace{0.04in} {\rm and} \hspace{0.04in}
\left(
\begin{array}{c}
\hat{{\bf x}}\\{\hat{\bf y}}\\{\hat{\bf z}}
\end{array}
\right)=
\textbf{\textsf{R}}^{\rm T}
\left(
\begin{array}{c}
\hat{{\bf x}}_b\\{\hat{\bf y}}_b\\{\hat{\bf z}}_b
\end{array}
\right),
\label{transf1v}
\end{equation}
and
\begin{equation}
\textbf{\textsf{R}}=\left(
\begin{array}{ccc}
\cos\alpha\cos\psi&\cos\alpha\sin\psi&-\sin\alpha\\
-\sin\psi&\cos\psi&0\\
\sin\alpha\cos\psi&\sin\alpha\sin\psi&\cos\alpha
\end{array}
\right),
\label{transf2}
\end{equation}
with $\textbf{\textsf{R}}^{\rm T}$ being the transpose of $\textbf{\textsf{R}}$. The corresponding unit vectors for radial, polar and azimuthal in spherical coordinates are signified by ${\hat{\bf r}}, {\hat\btheta},{\hat\bphi}$ and ${\hat{\bf r}}, {\hat\btheta}_b,{\hat\bphi}_b$, where the transformation is given by
\begin{equation}
\left(
\begin{array}{c}
\hat{{\bf r}}\\{\hat\btheta}\\{\hat\bphi}
\end{array}
\right)=
\textbf{\textsf{P}}
\left(
\begin{array}{c}
\hat{{\bf x}}\\{\hat{\bf y}}\\{\hat{\bf z}}
\end{array}
\right),
\label{transf6}
\end{equation}
and
\begin{equation}
\textbf{\textsf{P}}=\left(
\begin{array}{ccc}
\sin\theta\cos\phi&\sin\theta\sin\phi&\cos\theta\\
\cos\theta\cos\phi&\cos\theta\sin\phi&-\sin\theta\\
-\sin\phi&\cos\phi&0
\end{array}
\right)
\label{transf7}
\end{equation}
represents the transformation matrix.

Visibility of the pulsar emission is based on an idealized model in which radiation occurs at the source point that locates only within the open-field region \citep{Cordes78, KG03}, and the radiation is directed tangential to the local magnetic field line of dipolar structure \citep{Hibschman2001} and parallel to the line-of-sight direction \citep{YM14}. We assume emission from the highly relativistic particles is strongly confined to a narrow forward cone, i.e., the size of the forward cone is zero, and the aberration effect due to streaming along curved magnetic field lines is ignored. Then, the location of the visible point at a particular $\psi$ for known $\zeta$ and $\alpha$ is given by {\citep{Gangadhara04, YM14}}
\begin{eqnarray}
\cos 2\theta_{b{\rm V}} = \frac{1}{3} \Big(\cos\Gamma \sqrt{8+\cos^2\Gamma} - \sin^2\Gamma \Big), & \nonumber \\
\tan\phi_{b{\rm V}} = \frac{\sin\zeta\sin\psi}{\sin\alpha\cos\zeta-\cos\alpha\sin\zeta\cos\psi},
\label{eq:EmissionPtPhiB}
\end{eqnarray}
in the magnetic frame, where $\cos\Gamma=\cos\alpha\cos\zeta+\sin\alpha\sin\zeta\cos(\phi-\psi)$ is the half opening angle of the emission beam, or $(\theta_{\rm V}, \phi_{\rm V})$ in the osberver's frame using 
\begin{eqnarray}
\cos\theta&=&\cos\alpha\cos\theta_b-\sin\alpha\sin\theta_b\cos\phi_b, \label{eq:theta}\\
\tan(\phi-\psi) &=& \frac{\sin\theta_b\sin\phi_b}{\cos\alpha\sin\theta_b\cos\phi_b + \sin\alpha\cos\theta_b}. \label{eq:phi}
\end{eqnarray}  
The solutions relevant to our discussion correspond to the path traced by the visible point, referred to as the trajectory of the visible point, from the nearer of the two magnetic poles relative to $\psi =0$ where the impact parameter, $\beta =\zeta- \alpha$, is minimum. 

The motion of the visible point is periodic with the period of the star with components given by \citep{YM14}
\begin{equation} \label{eq:VelEmPtComp}
\omega_{{\rm V}\theta} = \omega_\star \frac{\partial\theta(\alpha,\psi)}{\partial\psi}, \quad \omega_{{\rm V}\phi} = \omega_\star \frac{\partial\phi(\alpha,\psi)}{\partial\psi}.
\end{equation}
Besides the case of $\zeta =0$, where $\omega_{\rm V} =\omega_\star$, $\omega_{\rm V}$ varies with $\psi$ and $\omega_{\rm V} < \omega_\star$ on the near side of the pulsar when the magnetic axis is around $\psi = 0$ and reaches the lowest speed at $\psi = 0$, but $\omega_{\rm V} > \omega_\star$ on the far side of the pulsar around $\psi=180^\circ$ with the maximum speed occurring at $\psi=180^\circ$. The overall average angular speed for one pulsar rotation is $\langle \omega_{\rm V} (\psi) \rangle = \omega_\star$. 

\end{document}